\documentclass[prc,aps,superscriptaddress,nofootinbib,twocolumn,noeprint]{revtex4-1}
\usepackage{amssymb}
\usepackage[dvips]{graphicx}
\usepackage[english]{babel}
\usepackage{indentfirst}
\usepackage{amsxtra}
\usepackage{amsmath}
\usepackage{supertabular}
\usepackage{multirow}
\usepackage[mathcal]{eucal}
\usepackage[usenames]{xcolor}
\usepackage{ulem}

\begin{document}
\title{Gamow-Teller strength in $^{48}$Ca and $^{78}$Ni with the charge-exchange
subtracted second random-phase approximation}
\author{D. Gambacurta}
\affiliation{INFN-LNS, Laboratori Nazionali del Sud, 95123 Catania, Italy}
\author{M. Grasso}
\affiliation{Universit\'e Paris-Saclay, CNRS/IN2P3, IJCLab, 91405 Orsay, France}
\author{J. Engel}
\affiliation{Department of Physics and Astronomy, CB 3255, University of Noth
Carolina, Chapel Hill, North Carolina 27599-3255, USA}

\begin{abstract}
We develop a fully self-consistent subtracted second random-phase approximation
for charge-exchange processes with Skyrme energy-density functionals.  As a
first application, we study Gamow-Teller excitations in the doubly-magic nucleus
$^{48}$Ca, the lightest double-$\beta$ emitter that could be used in an
experiment, and in $^{78}$Ni, the single-beta-decay rate of which is known.  The
amount of Gamow-Teller strength below 20 or 30 MeV is considerably smaller than
in other energy-density-functional calculations and agrees better with
experiment in $^{48}$Ca, as does the beta-decay rate in $^{78}$Ni.  These
important results, obtained without \textit{ad hoc} quenching factors, are due
to the presence of two-particle -- two-hole configurations.  Their density
progressively increases with excitation energy, leading to a long high-energy
tail in the spectrum, a fact that may have implications for the computation of
nuclear matrix elements for neutrinoless double-$\beta$ decay in the same
framework.  
\end{abstract}

\maketitle

Charge-exchange (CE) excitations \cite{oste,ichi} such as the Gamow-Teller (GT)
resonance are closely linked to electron capture and $\beta$ decay, which play
important roles in nuclear astrophysics \cite{langanke1,langanke2}.  They also
aid the construction of nuclear effective interactions, for which they constrain
couplings in the spin-isospin channel.  Finally, they are relevant to the
nuclear physics that affects neutrinoless double-$\beta$ ($0\nu\beta\beta$)
decay, in which two neutrons change into two protons
\cite{avignone2008,engel2017,mene}. The experimental observation of this rare
process would be a breakthrough for fundamental physics; it would mean that
neutrinos are Majorana particles and would imply new phenomena beyond the
Standard Model that could be related to the matter-antimatter asymmetry in the
universe.  The rate of $0\nu\beta\beta$ decay depends on nuclear matrix elements
that may only be determined theoretically.  At present, predictions for these
matrix elements differ from one model to the next by factors of two or three, an
amount that is too large to allow the efficient planning and interpretation of
experiments.  

Any model that hopes to describe double-$\beta$ decay must be able to predict
the distribution of GT strength because the GT operator, multiplied by the
axial-vector coupling constant $g_{A}$, is the leading contribution to the
operator that governs $\beta$ decay.  Recent work
\cite{shimizu,santopinto18,ferreira20} that correlates calculations of double-GT
and $0\nu\beta\beta$ matrix elements even suggests that one could deduce the
latter from double-CE experiments \cite{cappuzzello,lenske}.  Most
double--$\beta$ emitters that could be used in experiments are still too complex
to easily treat from first principles with \textit{ab-initio} methods (though
recently, such methods were applied to $^{48}$Ca \cite{yao,novario20,belley20},
and to $^{76}$Ge and $^{82}$Se \cite{belley20}).  More phenomenological
approaches are therefore still important, even necessary
\cite{SM,tokyo,michigan,ibm,nredf1,nredf2,redf1,redf2, tubingen,jyva,ch}.
However, these theoretical schemes do not correctly describe the available data
for GT excitations and $\beta$-decay half-lives and must resort to \textit{ad
hoc} ``quenching factors'' to obtain reasonable results for GT strength below 20
or 30 MeV of excitation energy.  In Ref.\ \cite{cao}, for example, models based
on the random-phase approximation (RPA) overestimate the strength significantly.
This kind of over-prediction is usually ascribed to missing physics, for example
the $\Delta$ excitation \cite{bohr81} or complex configurations such as
two-particle -- two-hole (2p2h) excitations \cite{bh,ait,klein85}.  The
results in Ref.\ \cite{cao} indicate that higher-order correlations are needed
beyond those in the RPA, which is essentially a time-dependent version of
mean-field theory.

\textit{Ab-initio} work with operators and currents from chiral effective field
theory has recently had some success in explaining the quenching in $\beta$
decay.  Reference \cite{gysbers} showed that correlations omitted from the shell
model and from mean-field-based calculations, together with two-body weak
currents, account for most of that quenching.  Reference \cite{ekstrom} showed
that the same effects quench the integrated $\beta$ strength function.  But weak
two-body currents play no obvious role in charge-exchange transitions, and so
the implications of this last result for our work are not clear.  Similarly,
starting from realistic potentials, the authors of Ref.\ \cite{coraggio} used
many-body perturbation theory to derive effective shell-model operators that
implicitly include correlations from outside shell-model spaces, obtaining the
correct quenching of GT strength in several nuclei of interest for
double-$\beta$ experiments, but failing to do so in the lightest of these,
$^{48}$Ca.   They also had difficulty in that nucleus with two-neutrino
double-$\beta$ decay, a very closely related process.  

Finally, EDF-based models that go beyond mean-field theory have been proposed
for CE excitations, for example in both relativistic (see Refs.
\cite{robin1,robin2} for the most recent developments) and nonrelativistic (see
for instance Ref.\  \cite{niu}) particle-vibration-coupling models.  The
predicted integrated strengths are always better than in the (Q)RPA. Again,
however, the improvement is minor for $^{48}$Ca.  Reference \cite{niu} shows
that the GT strength below 20 MeV continues to be significantly overestimated in
that nucleus, even when beyond-mean-field correlations are included.  In this
paper a better description of the GT$^-$ strength --- measured in
charge-exchange reactions by adding a proton and removing a neutron --- in
$^{48}$Ca, and of the GT $\beta$ decay of $^{78}$Ni, are achieved with a
subtracted second RPA (SSRPA).  
    
Many-body theorists employ RPA-based schemes extensively in atomic,
solid-state, and nuclear physics, as well as in quantum chemistry.  Extensions
are useful wherever beyond-mean-field correlations play an important role.
Second RPA, which includes 2p2h configurations for a richer description of the
fragmentation and widths of excited states, has thus made its way from nuclear
physics, where it was born, to mesoscopic physics \cite{gamba09} and chemistry
\cite{peng14}.

Versions of the RPA for CE processes were introduced several decades ago;
references \cite{auerbach81,auerbach82,lane80} contain useful discussions of the stability of
the Hartree-Fock solution with respect to isospin excitations.  The non--trivial
step of constructing a full CE second RPA was taken in later, in Ref.\
\cite{ait}.  There, one can find expressions for the Hamiltonian matrix, which
contains a 1p1h sector characterized by the matrices called $A_{11}$ and
$B_{11}$, a sector that mixes 1p1h and 2p2h configurations, with the matrices
$A_{12}$ and $B_{12}$, and a pure 2p2h sector, with the matrices $A_{22}$ and
$B_{22}$. Because the diagonalization of large dense matrices was impossible at
that time, the Hamiltonian matrix was drastically simplified by neglecting the
interaction among 2p2h configurations. That step allowed the full SRPA
diagonalization to be replaced by an RPA-type computation with an
energy-dependent Hamiltonian.  Much more recently, a self-energy subtraction
procedure was designed for extensions of RPA \cite{tse2013} and implemented in
charge-conserving second RPA
\cite{gamba2015,PLB,vasseur2018,grasso2018,gamba2019,grasso2020} to make the
treatment of excitations consistent with ground-state density-functional theory,
guarantee Thouless stability, and eliminate ultra-violet divergences.  

The CE second RPA developed here is the first that is fully self consistent and
includes a subtraction procedure that, just as in the charge-conserving case,
corrects the response to make it consistent with ground-state density-functional
theory at zero frequency.  We apply it in together with the Skyrme interaction
\cite{sk1,sk2,vautherin} SGII \cite{SGII,SGII-2}.  The subtraction procedure
requires the inversion of the large matrix $A_{22}$.  To make the problem
tractable, we consistently cut off 2p2h configurations at 40 MeV, both in the
diagonalization of $A_{22}$ and its inversion, having verified that results do
not change significantly when the cutoff is raised beyond that level.   

GT strength is constrained by the Ikeda sum rule, which relates the integrated
strengths $S$ to the number of neutrons $N$ and protons $Z$ in the nucleus: 
\begin{equation}
S_{GT^-}-S_{GT^+}=3(N-Z).
\label{ikeda}
\end{equation}
The sum rule is model independent under the condition of completeness of states
and given the properties of isospin operators. It holds in RPA approaches and
their extensions if, as in this work, the quasiboson approximation applies.  

In nuclei with a significant neutron excess, such as $^{48}$Ca and $^{78}$Ni,
the GT$^-$ strength is much larger than GT$^+$ strength, measured by adding a
neutron and removing a proton.  The excitation operator for GT$^-$ transitions
can be written as
\begin{equation}
\hat{O}^-=\sum_{i=1}^{A} \sum_{\mu} \sigma_{\mu}(i) \tau^-(i)
\end{equation} 
where $A$ is the number of nucleons, $\tau(i)^-$ is the isospin-lowering
operator $\tau^- \equiv t_x-it_y$ for the $i^{th}$ nucleon, and
$\sigma_{\mu}(i)$ is the corresponding spin operator.  Because $S_{GT^{-}} $ is
so much larger than $S_{GT^{+}}$ the Ikeda sum rule is essentially a measure of
the total GT$^{-}$ strength (as we will see later numerically).

We begin with the case of $^{48}$Ca.   Reference \cite{yako} reports the
results of a $^{48}$Ca(p, n) and $^{48}$Ti(n, p) experiments at a beam energy of
300 MeV at the Research Center for Nuclear Physics in Osaka.  The total GT$^-$
strength below 30 MeV (which probably includes some contributions from isovector
spin-monopole excitations) is only 64 $\pm$ 9\% of that given by the Ikeda sum
rule.  The location of this ``missing strength'' has long been a mystery for
nuclear physics.

\begin{figure}
\includegraphics[width=\columnwidth]{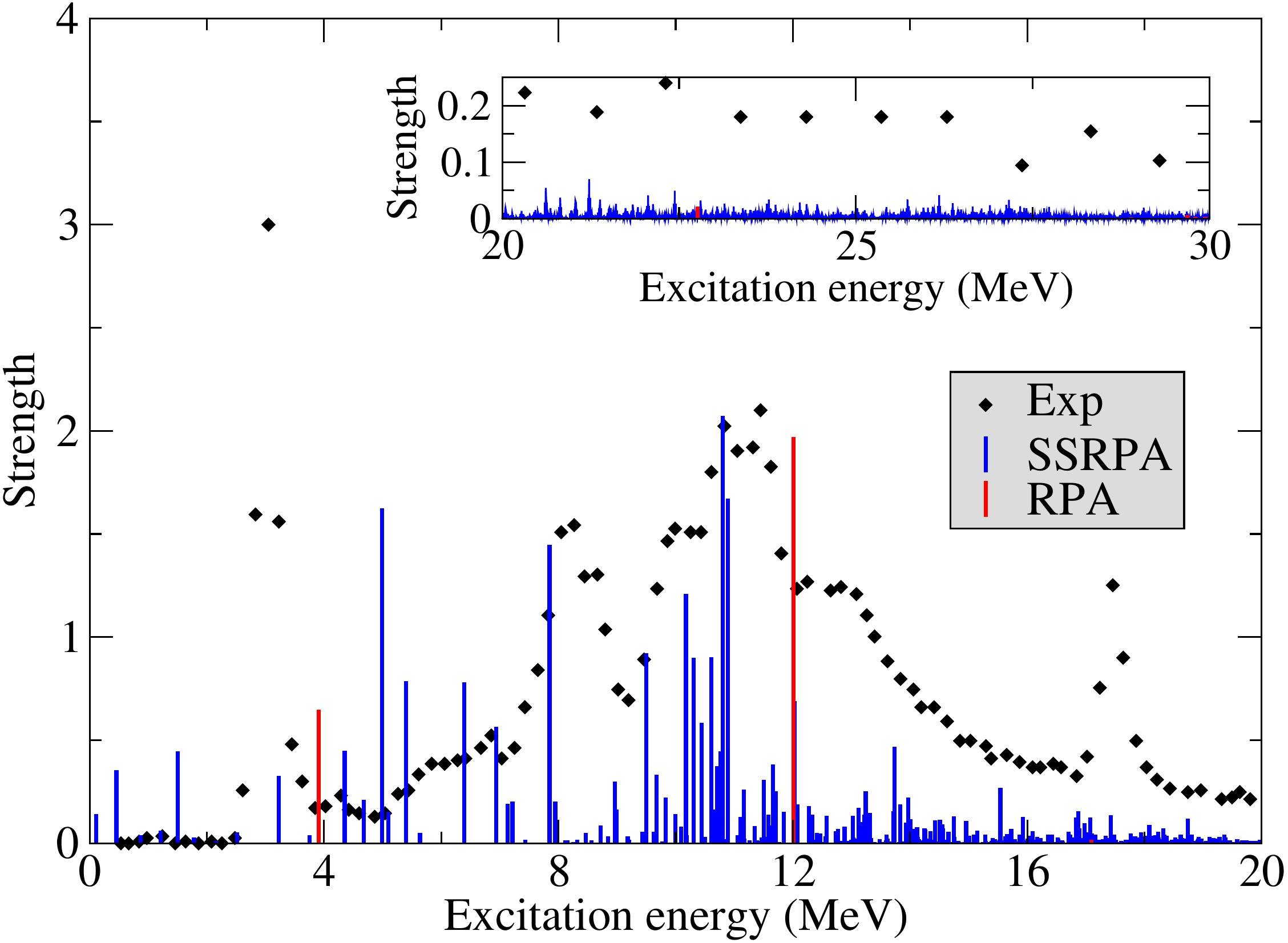}
\caption{Experimental GT$^-$ plus isovector spin-monopole strength in MeV$^{-1}$
\cite{yako} and discrete RPA and SSRPA strength distributions (no units)
obtained with the Skyrme parameterization SGII, for $^{48}$Ca .  The RPA
strength has been divided by nine and the SSRPA strength by two so that the
discrete distributions can be displayed on the same figure as the continuous
experimental distribution (see text).  The insert shows the energy region
between 20 and 30 MeV.} \label{spec}
\end{figure}

Compared to other EDF approaches (both mean-field and beyond-mean-field) ours
better predicts the strength distributions, so that we obtain a much more
accurate value for the sum of the strength up to 20 or 30 MeV, without resorting
to quenching factors.  The 2p2h configurations, which increase in density with
excitation energy, lead to a long high-energy tail that draws strength from
lower energies.  The ``missing strength'' is thus spread out over a large range
at higher energies, making it hard to discriminate from background.

Figure \ref{spec} shows the experimental strength extracted from Ref.\
\cite{yako}. One of the most important features of the SSRPA is its ability to
describe the width and fragmentation of excitation spectra. This asset is
visible in the figure, which contains both the RPA and SSRPA discrete-strength
distributions.  Because the experimental strength is a continuous function of
energy, it has different units from the discrete theoretical strengths, and the
absolute strength values are thus not comparable.  However, by plotting the
discrete spectra one can compare the location and fragmentation of the main
peaks, without generating any artificial spreading by folding. To better display
the results in the figure, we rescale the RPA and SSRPA discrete strengths so
that their respective highest peaks have approximately the same height as the
corresponding experimental peak.  To achieve this, we divide the RPA strength by
nine and multiply the SSRPA strength by two.   

The SSRPA strength is indeed quite fragmented, particularly in the region
between 6 and 16 MeV, where three groups of peaks are concentrated around 8, 11,
and 14 MeV, in accordance with the experimental distribution of peaks.  One may
also observe another group of much weaker peaks concentrated around 17 MeV,
which corresponds to the location of the highest-energy experimental peak.
Finally, a very dense high-energy SSRPA tail is visible in the insert, which
focuses on the energy region between 20 and 30 MeV. Such a tail is completely
absent from the RPA spectrum, which is composed of a few well separated peaks
and misses the complex structure of the experimental strength.  The long
high-energy tail is indeed the explanation for the missing strength at lower
energies.

\begin{figure}
\includegraphics[width=\columnwidth]{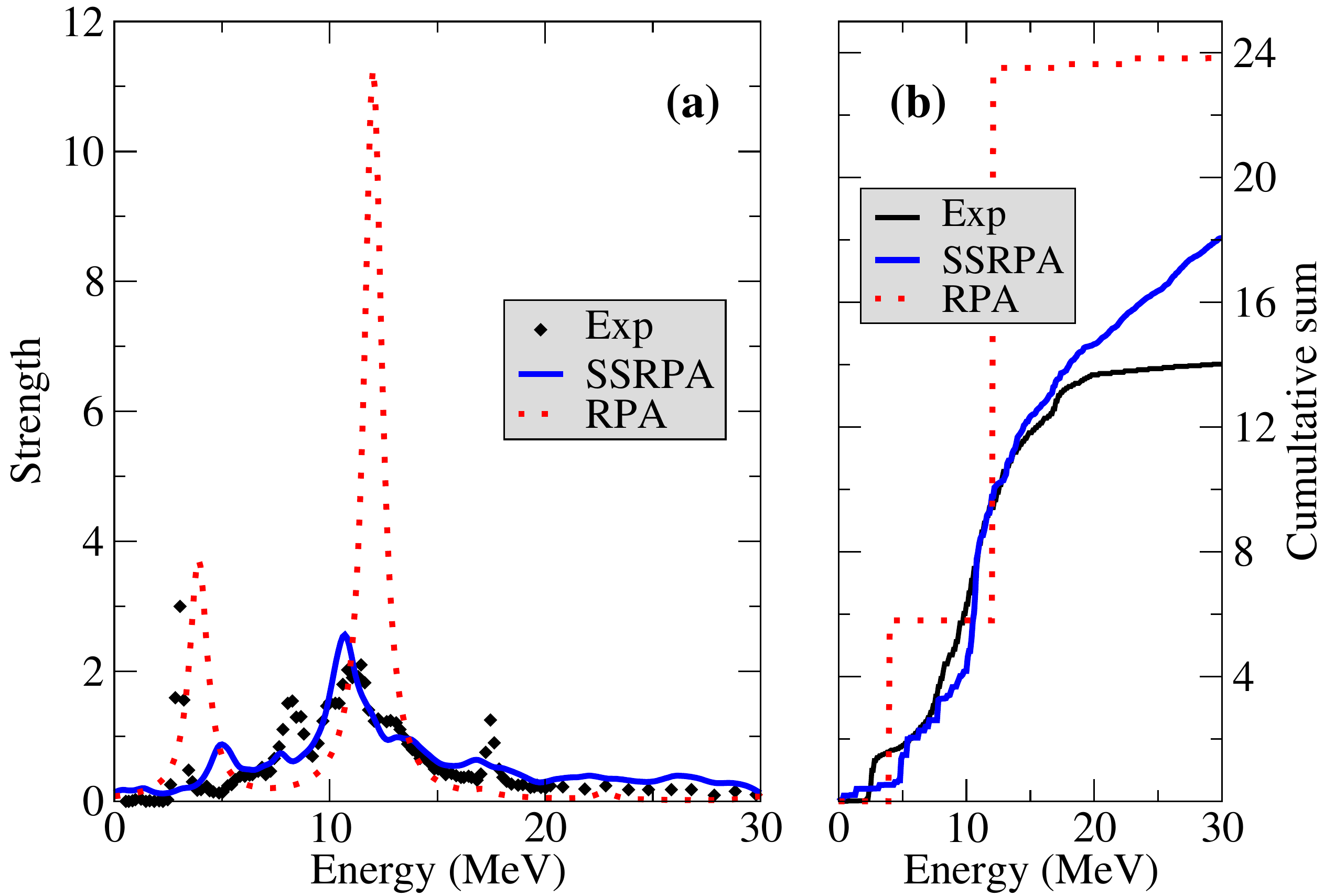}
\caption{(a) GT$^-$ strength distributions in MeV$^{-1}$. The experimental
points are extracted from Ref.\ \cite{yako}. The RPA and SSRPA responses,
computed with the parameterization SGII, are folded with a Lorentzian having a
width of 1 MeV. (b) Cumulative strengths up to 30 MeV.} 
\label{folded}
\end{figure}

The very lowest-energy part of the SSRPA spectrum is less satisfactory than the
rest, with a main peak predicted at about 5 MeV; the lowest experimental peak,
by contrast, is located at 3 MeV.  The SSRPA does predict some fragmented
strength is in the region around 3 MeV, however.  The RPA discrete spectrum in
Fig.\ \ref{spec}, by contrast, shows only a single visible peak at 4 MeV.

To more directly compare the theoretical and experimental strengths, we have
folded our response functions together with a Lorentzian distribution of width 1
MeV.  Panel (a) of Fig.\ \ref{folded} presents the folded RPA and SSRPA strength
distributions along with the experimental distribution. Panel (b) shows the
cumulative strength as a function of energy up to 30 MeV.  As we have already
seen in the discrete spectra, the SSRPA reproduces the GT distribution quite
well, with the exception of the lowest-energy peak.  The RPA, on the other hand,
cannot reproduce the complex structure of the spectrum.  

The most striking result is in panel (b). The SSRPA cumulative strength is
greatly reduced from that of the RPA and is in much better agreement with the
experimental value, even at low energies.  Furthermore, the SSRPA curve is
smooth and follows the experimental profile (owing to the physical description
of widths and fragmentation, and to the subtraction procedure, which is needed
to place centroids at the correct energies), except beyond 20 MeV, where the
tail is a little too high, The RPA curve, by contrast, shows steps because of
its very few well separated peaks.  The improvement with respect to the RPA is
more significant than in other beyond-mean-field approaches.  In Ref.\
\cite{niu}, for example, the same Skyrme interaction SGII produces more than 20
units of strength below 20 MeV.

The ratio between the experimental and the theoretical integrated strength below
20 MeV is 0.58 for the RPA and 0.93 for the SSRPA, showing that quenching
factors are not needed in SSRPA.  In the particle--vibration--coupling
calculations of Ref.\ \cite{niu}, this ratio is $\le$ 0.68, with the upper limit
corresponding to an integrated strength of 20 units.  The explicit inclusion of
2p2h configurations efficiently generates the high-energy tail that accounts for
the missing strength.  When we integrate the strength up to 70 Mev, we obtain a
total of 23.84.  The integrated $\beta^{+}$ strength at that energy is only
0.10, so we reproduce the Ikeda sum of 24 to within about 1\%. 

\begin{figure}[b]
\includegraphics[width=\columnwidth]{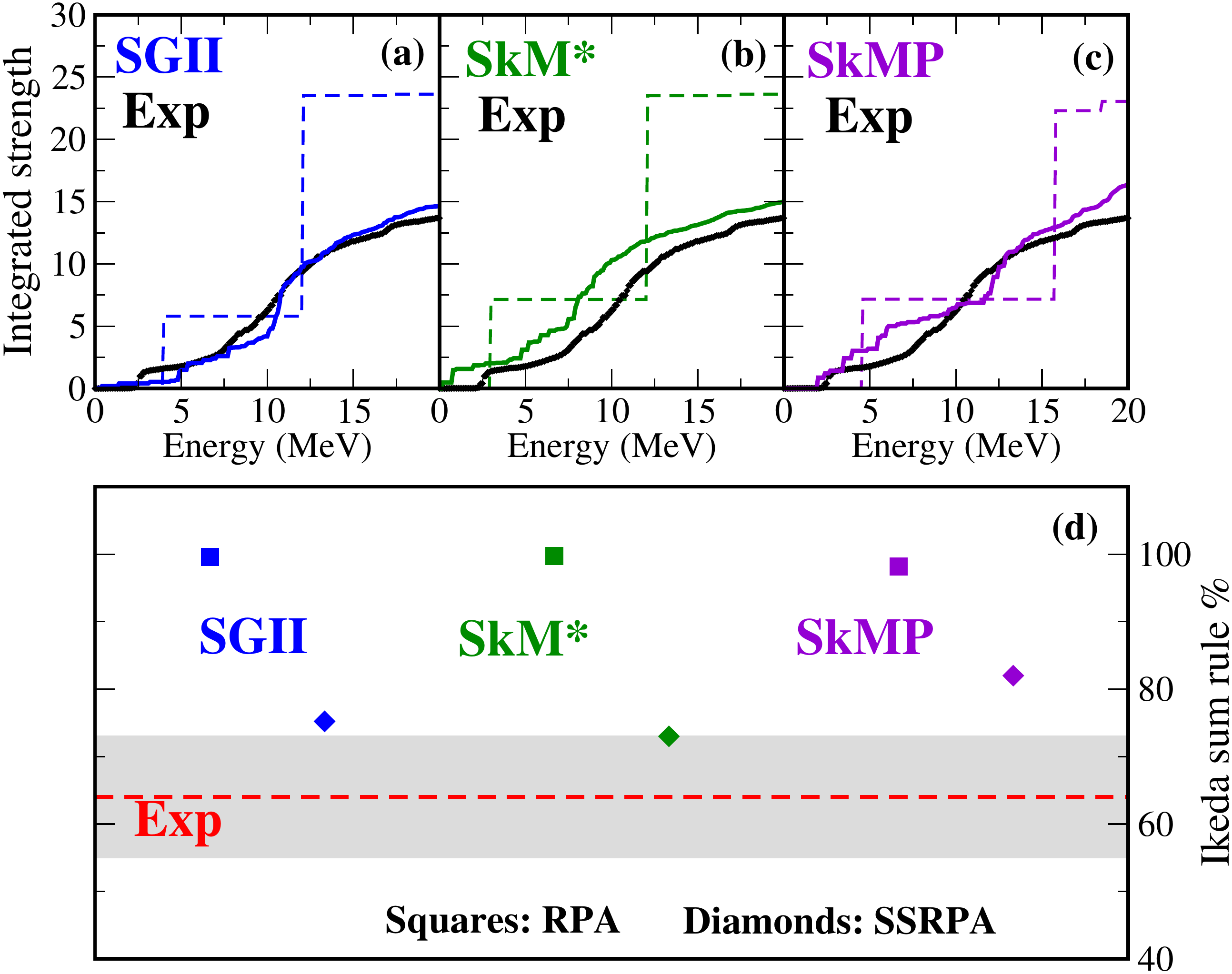}
\caption{(a), (b), (c) Strengths integrated up to 20 MeV. The black symbols
represent the experimental values \cite{yako}. The solid (dashed) lines
correspond to the SSRPA (RPA) results obtained with the parameterizations SGII
(a), SkM$^*$ (b), and SkMP (c). (d) Experimental percentage of the Ikeda sum
rule below 30 MeV extracted from Ref.\ \cite{yako} (red dashed horizontal line),
and its associated uncertainty (grey area). RPA and SSRPA percentages obtained
with the parameterizations SGII, SkM$^*$, and SkMP are also shown.}
\label{percentuale}
\end{figure}

To generalize our analysis, and to show that the decrease in strength below 20
MeV is an intrinsic effect of the SSRPA and not an artifact of a specific
parameterization, we show in Fig.\ \ref{percentuale} RPA and SSRPA results
obtained with three different parameterizations --- SGII, SkM$^*$ \cite{bartel},
and SkMP \cite{hen} --- together with their experimental counterparts.  Panels
(a), (b), and (c) show the cumulative strengths up to 20 MeV, whereas panel (d)
displays the percentages of the Ikeda sum rule from strength below 30 MeV.  For
this last quantity, Ref.\ \cite{yako} has reported both the experimental value
and its uncertainty.  In all three upper panels the integrated strength in the
SSRPA is smaller than in the RPA, especially for SGII and SkM$^*$.  With these
two parameterizations, the percentages of the Ikeda sum in panel (d) are quite
close to the experimental value. The upper panels show that the detailed
structure of the experimental spectrum is best reproduced by the
parameterization SGII; in panel (a) the curve follows the experimental profile
very closely. 

\begin{figure}
\includegraphics[width=\columnwidth]{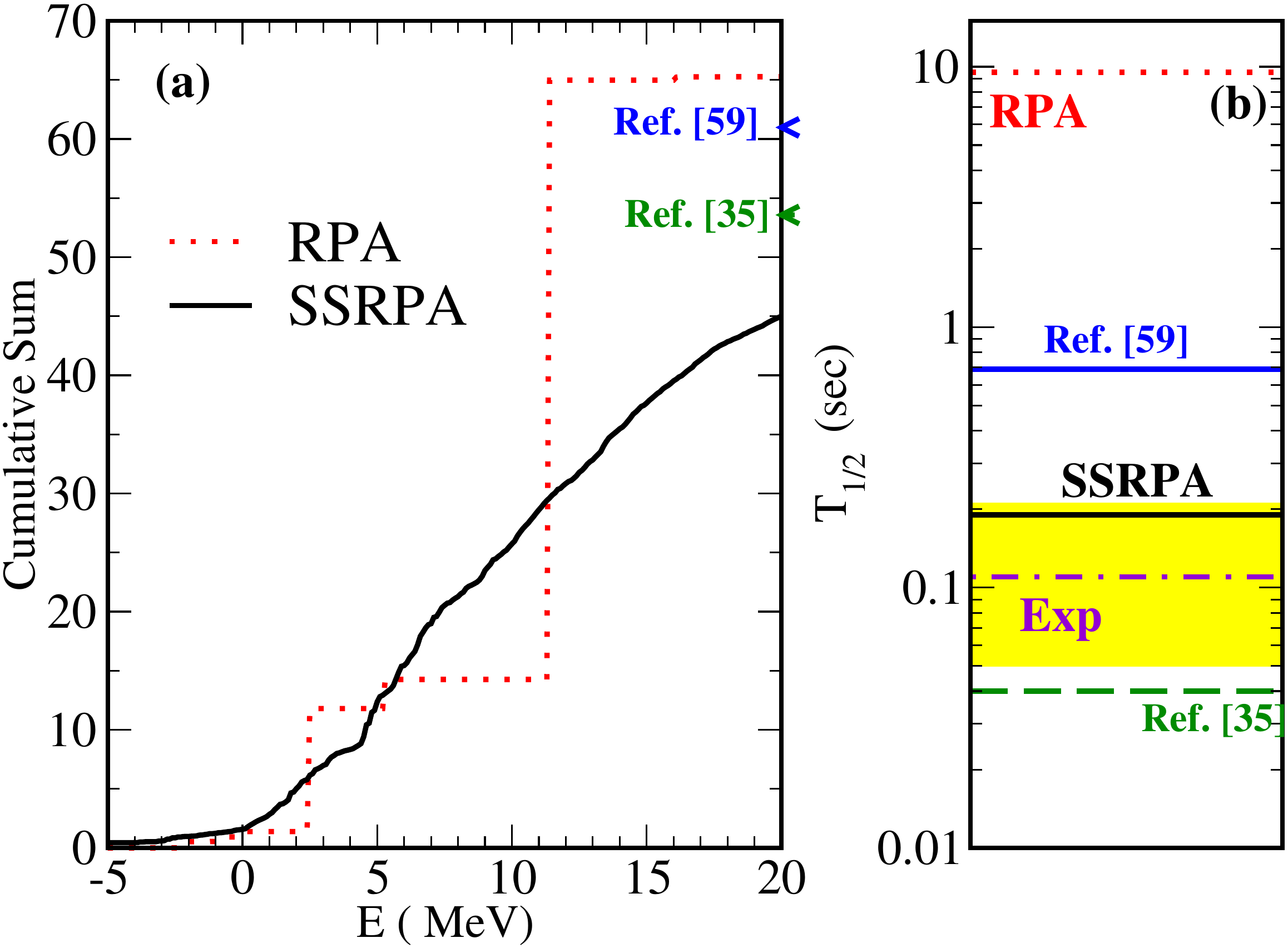}
\caption{(a) Cumulative sum for different models (see legend and text) for the
nucleus $^{78}$Ni; (b) $\beta$-decay half-life for $^{78}$Ni predicted by SSRPA,
compared with predictions of other models and the experimental value
\cite{hosmer05}. The yellow band represents the experimental uncertainty.}
\label{beta}
\end{figure}

We turn now to $^{78}$Ni.  Though the GT spectrum of this nucleus has never
been measured, its $\beta$-decay lifetime is known \cite{hosmer05}.  Our GT$^{-}$
strength, integrated up to 20 MeV, appears in Fig.\ \ref{beta}(a) from both the
RPA and SSRPA (this time in a diagonal approximation so that we can invert the
larger $A_{22}$ in this heavier nucleus).  The figure also displays two other
theoretical values for the integrated strength at 20 MeV, from the
beyond-mean-field calculations of Refs.\ \cite{robin1} and \cite{niu15} (for the
latter, with the same interaction SGII as we use).  Our value is significantly
lower than those of both the RPA and the other two beyond-mean-field
calculations. 

Finally, we use our computed GT strength to obtain the $\beta$-decay half-life
of $^{78}$Ni.  Because correlations make {\it{ad-hoc}} quenching unnecessary in
our approach, we use the bare value 1.28 for $g_{A}$, the weak axial-vector
coupling constant.  Our result appears in Fig.\ \ref{beta}(b), together with the
experimental value \cite{hosmer05}.  The SSRPA half-life is 0.19 seconds, very
close to the experimental half-life.  The RPA half-life, by contrast, is 9.51
seconds with the bare value of $g_{A}$.  The predictions of Refs. \cite{robin1}
(in the most complete scheme that includes polarization effects related to CE
phonons) and \cite{niu15} (with the interaction SGII) are 0.04 and 0.69 seconds
respectively, again without $g_{A}$ renormalization.  Both these results are
farther from the experimental half-life than ours.

In summary, we have presented a crucial improvement in the description of the GT
strength for the lightest double-$\beta$ emitter, the nucleus $^{48}$Ca, and the
heavier nucleus $^{78}$Ni. This achievement was made possible by the
beyond-mean-field EDF-based CE-SSRPA, developed here for the first time. We have
used the approach to compute GT$^-$ strengths with the Skyrme interaction SGII,
and shown that it reproduces the complex fragmented spectrum in $^{48}$Ca much
better than does the RPA.  Our most important result is that the total SSRPA
strength below 20-30 MeV is much smaller than in other mean-field and
beyond-mean-field EDF models, and in better agreement with the corresponding
experimental values, without the use of \textit{ad hoc} quenching factors.  By
working with two additional Skyrme parameterizations we showed that these
successes are due primarily to our many-body method, the key ingredient of which
is the explicit inclusion of 2p2h configurations. Their density strongly
increases with the excitation energy, leading to a high-energy tail in the
spectrum.  We showed that the same effects reduce the GT strength in the nucleus
$^{78}$Ni, so that the predicted $\beta$-decay half-life agrees better with
experiment than do other beyond-mean-field models.

The ability to describe CE strength may have a strong impact in astrophysical
scenarios where GT resonances play an important role.  It also promises to
improve EDF-based calculations of the nuclear matrix elements governing
$0\nu\beta\beta$ decay, a process at the intersection of several scientific
domains.  Our less accurate reproduction of the weak low-energy part of the
$^{48}$Ca spectrum is unlikely to have a large effect on the $0\nu\beta\beta$
matrix element, which contains an unweighted sum over states at all energies.
We plan to apply our approach to open--shell nuclei by including pairing
correlations of both the usual isovector type and the isoscalar type that are
important for $\beta$ and double-$\beta$ decay \cite{ch,mu2014,bai2014,niu2016}.  

\begin{acknowledgments}
M.G.\ acknowledges funding from the European Union Horizon 2020 research and
innovation program under Grant No. 654002 and from the IN2P3-CNRS BRIDGE-EDF
project.  J.E.\ acknowledges support from the U.S.\ Department of Energy (DOE),
Office of Science, under Grant No.\ DE-FG02-97ER41019.
\end{acknowledgments}


%

\end{document}